# Nitrogen backbone oligomers

[1,2]H. Wang, [1]M. I. Eremets*, [1,3]I. Troyan, [2]H. Liu, [2]Y. Ma[+], [1]L. Vereecken

[1]Max Planck Institute for Chemistry, Biogeochemistry Department, PO Box 3060, 55020 Mainz, Germany

[2]State Key Lab of Superhard Materials, Jilin University, Changchun 130012, P. R. China

[3]Institute of Crystallography, Russian Academy of Sciences, Leninsky pr. 59, Moscow 119333, Russia

*e-mail: m.eremets@mpic.de

[+]mym@jlu.edu.cn

In contrast to carbon, which forms long polymeric chains, all-nitrogen chains are very unstable. Here we found that nitrogen and hydrogen directly react at room temperature and pressures of ~35 GPa forming a mixture of nitrogen backbone oligomers – chains of single-bonded nitrogen atom with the rest of the bonds terminated with hydrogen atoms – as identified by IR absorption, Raman, X-ray diffraction experiments and theoretical calculations. The pressure required for the synthesis strongly decreases with temperature to ~ 20 GPa at 550 K. At releasing pressures below ~10 GPa, the product transforms into hydrazine. Our findings might open a way for the practical synthesis of these extremely high energetic materials as the formation of nitrogen-hydrogen compounds is favorable already at pressures above 2 GPa according to the calculations.

Nitrogen is a unique element in terms of its huge difference in binding energy between nitrogen atoms attached with single (or double) bonds, versus the very strong, short triple bond N≡N. This latter bond (−477 kJ mol$^{-1}$) is one of the strongest chemical bonds known, while the N−N bond is much weaker (−80 kJ mol$^{-1}$) (1). Because of this, many nitrogen compounds are high energy density materials (HEDM) that release a large amount of energy at their, possibly explosive, decomposition to the most stable species – dinitrogen molecules (2). All-nitrogen crystal, the so-called "polymeric nitrogen," has been formed at very high pressures as a three-dimensional crystalline (1, 3) or as a disordered network of single-bonded nitrogen atoms (4). Chains of nitrogen atoms were also predicted (5-7). Polymeric nitrogen is considered a material that can store an ultimately large amount of chemical energy. Unfortunately, megabar pressures (100 GPa) are needed to synthesize all-nitrogen polymers. This precludes any practical applications which should be in the pressure range of ~1 GPa or ideally ~0.1 GPa – the range of synthesis of ammonia in the Haber-Bosch process.

Another route for nitrogen high energy density materials is formation of nitrogen-hydrogen compounds. However, the weak single and double bonds prevent easy formation of large molecules or polymers. Recent work has discovered molecules with a larger number of nitrogen atoms in a row: N5 (8), N8 (9), and N10 (10); however, these compounds are either ionic or stabilized by carbon. Larger metastable polynitrogen molecules were predicted (11) but not found experimentally yet.

Application of very high pressures dramatically changes the chemistry of nitrogen, where single covalent bonding becomes preferable. In the present work we explored a synthesis of single-bonded nitrogen-hydrogen compounds under high pressure with an idea that addition of hydrogen might reduce the pressure needed for synthesis of new energetic materials, and make these more stable in comparison to the purely nitrogen polymers. Hydronitrogen polymer was predicted by recent evolutionary *ab initio* calculations (12): at pressures > 36 GPa NH chains were formed from the ammonium azide precursor. Experimentally these calculations seem not to be supported as ammonium azide is stable at pressures of at least ≈55 GPa at room temperature (13), probably even higher pressure is required for the polymerization. In principle, diimide $(NH)_2$ in which nitrogen atoms are double bonded might serve as a monomer but it is very unstable and difficult to handle. Finally, hydrazine does not polymerize as we observed no qualitative changes in the IR and Raman spectra by increasing the pressure up to 60 GPa and subsequently releasing the pressure (Fig. S1).

Because of the apparent lack of nitrogen-hydrogen precursors we explored a direct reaction using molecular nitrogen and hydrogen as the starting materials. Only a few studies on hydrogen-nitrogen mixtures under pressure are available. Changes in the Raman spectrum shows that hydrogen incorporates into the nitrogen lattice and interacts with the neighboring nitrogen atoms (14-16). An inclusion compound $(N_2)_{12}D_2$ was observed from a 1:9 $D_2:N_2$ mixture(16). Ciezak et al (15) studied a 1:2 $H_2:N_2$ mixture at room temperature and found an indication for chemical transformation as new bands in Raman spectra appeared, which were assigned to N-N bending and stretching. On the other hand, molecular nitrogen and hydrogen vibrons were observed even at the highest pressures of 85 GPa, such that no definite conclusion was drawn on the behavior of this mixture (15).

In this work, we study $H_2/N_2$ mixtures at high pressures experimentally and theoretically, aiming for evidence of the formation of nitrogen-hydrogen compounds, and identifying the compounds formed.

**Experimental Results**

We systematically studied $N_2/H_2$ mixtures with a wide range of nitrogen concentration ratios (5%, 10%, 20%, 50%, and 80%) as well as pure nitrogen and hydrogen at pressures of up to 70 GPa. For all mixtures we obtained consistent results and will illustrate results taken for different $N_2/H_2$ compositions. For instance, for the 1:9 $N_2/H_2$ mixture, the Raman spectra deviated from those of pure nitrogen and hydrogen at pressures as low as ~10 GPa: the nitrogen vibron peak split (Fig. 1a), and two strong satellites of the hydrogen vibron emerged (Fig. 1c) indicating that hydrogen and nitrogen are mixed and strongly interact with each other. At these pressures, however, they remained in the molecular state because vibron excitations in the Raman spectra persisted.

A major change occurs above ~35 GPa: the intensity of $H_2$ and $N_2$ Raman vibrons strongly decreases and then even disappears at $\cong$ 50 GPa (Fig. 1a–d, Figs. S2, S3). This transformation is accompanied by a large decrease in volume, as indicated by the decrease in the frequency of the vibrons and the substantial sharpening of the high-frequency edge of the Raman signal from the diamond anvil close to the sample (Fig. S3c). The new phase is amorphous because the X-ray diffraction pattern disappears at pressures above ~50 GPa (Fig. S2c,d).

The pressure needed to induce formation of the new phase strongly reduces with temperature as we found by studying a 1:1 $N_2:H_2$ mixture. Instead of pressure as was done at room temperature, we changed temperature at certain pressure. The transformation was monitored on the basis of the disappearance of the hydrogen vibron. The IR absorption spectra were recorded after cooling to room temperature to verify the transformation. Heating the sample at 44 GPa, for instance, shows a decline in the $H_2$ vibron starting at 360 K, and the $H_2$ signal vanishes above 400 K. At 20 GPa, the hydrogen vibron disappeared at 550 K. The pressure of the transformation drops with temperature approximately linearly (Fig. 2). Owing to hydrogen diffusion out of the sample through the metallic gasket, we were unable to achieve higher temperatures.

IR spectroscopy provided us with several clues as to nature of the new phase. First of all, the strong absorption band at 3300 $cm^{-1}$ (Fig. 3a) is characteristic of N–H vibrational stretching modes, indicating that the high-pressure transformation involves a chemical reaction between nitrogen and hydrogen. Second, upon releasing the pressure below ~10 GPa the product sharply transforms into hydrazine as it is unambiguously identified by the IR and Raman spectra (Figs. 3b, 4 (17). This transformation indicates that the product relates to hydrazine. However, the absence of torsion mode at ~600 $cm^{-1}$ and wagging mode at ~1300 $cm^{-1}$ (Fig. S4) indicates that longer molecules than hydrazine are synthesized. However, precise identification of such a disordered chain-length material with spectroscopic and X-ray data alone cannot be done because the product is apparently is a disordered mixture of different molecules that results in the broad IR and Raman spectra (Figs 3,4) and the diffuse X-ray.

**Theoretical calculations**

To gain further insight in the possible nature of the high-pressure amorphous phase formed in the hydrogen-nitrogen mixture, we performed theoretical calculations based on two different approaches: metadynamics calculations (18, 19) to simulate structural transformations in the mixture and quantum chemical calculations to examine the energetics and spectrum of hydrogen-nitrogen molecules.

Metadynamics simulations were performed at 300 K for a 1:4 ratio of $N_2$:$H_2$ in an initial mixed molecular arrangement (Fig. 5a). At 30 GPa no obvious structural changes or reactions were observed after 100 metasteps, which is typically long enough to model phase transitions, in good agreement with our experimental results. Upon compression to 60 GPa, reaction between nitrogen and hydrogen was found, where four kinds of hydronitrogen compounds were stabilized: $H_2N-NH_2$ (hydrazine), $H_2N-NH-NH-NH_2$ (N4-II), $H_2N-N=N-NH_2$ (labeled N4-I), and $H_2N-NH-N(NH_2)-NH-NH_2$ (N6), in ratios of 7:3:1:1. Their formation is associated with a large enthalpy gain, about -298 kJ mol$^{-1}$ per $N_2H_8$ formula unit (corresponding to the 1:4 $N_2$:$H_2$ mixture) relative to the reference $N_2$ ($R\bar{3}c$ structure) and $H_2$ ($P6_3/m$) (Fig. 5c). The formation of nitrogen-hydrogen compounds is favorable already at pressures above 2 GPa. However there is a large kinetic barrier for the polymerization while it is quite difficult to estimate its value. We have calculated the electronic band structure and density of states of N-H compounds at 60 GPa. The results showed that the formed N-H compounds are insulating with a band gap of ~3.7 eV (Fig. S5) rather than a metal predicted for a chain consisted only of nitrogen atoms (5-7). Higher temperatures promote the polymerization: at 500K, longer chains are created in larger amounts (Fig. S6).

Our quantum chemical calculations are based at the M06-2X/aug-cc-pVDZ level of theory (see Methods and the full description in the SI). We find that the azanes, the systematic class of $N_nH_{n+2}$ compounds analogous to the carbon-based alkanes, are energetically stable compounds even at zero pressure, in a qualitative agreement with the above *ab initio* calculations (Fig. 6). They have well-defined minima on the potential energy surface, and therefore can exist as separate compounds, even if their thermal lifetime is currently unknown. This also agrees with literature data (20). Linear, unbranched polymeric chains, $NH_2-(NH)_n-NH_2$, were studied for up to N10. Contrary to hydrocarbons which have a linear backbone, the most stable conformer of the azanes is a spiral-shaped nitrogen backbone (Fig. 6), owing to the interactions between the free electron pair and the bonds on the nitrogen atoms. Their formation enthalpy, relative to $N_2$ and $H_2$, increases by about 84 kJ mol$^{-1}$ for every additional NH monomer, making azanes highly energetic materials with an energy density comparable to polymeric nitrogen.

The longer azanes were found to be thermally less stable than hydrazine. Thermal dissociation of hydrazine requires over 250 kJ mol$^{-1}$, forming two $^\bullet NH_2$ radicals. For larger chains, >N4, breaking of the internal N-N bonds requires only 125 kJ mol$^{-1}$, leading to the more stable $-N^\bullet H$ radicals. This could indicate that longer azanes have a higher tendency to decompose to smaller compounds, even though the predicted barrier still affords long thermal lifetimes at 300 K. The formation, destruction and transformation of azanes appear to involve a complex set of radical reactions, which may be at work in the experimental setup. The full quantum chemical characterization of this chemistry is beyond the scope of this work.

**Discussions**

Combining the experimental data with the theoretical predictions leads to further insights in the possible structure of the amorphous phase formed at high pressures. The metadynamics calculations show that NH-chains longer than hydrazine can be formed. The quantum chemical calculations on the formation enthalpies indicate that, per NH-unit, the formation enthalpies are very similar across longer azane chains. This suggests that, if hydrazine can be formed, longer azanes can be formed as well, without a prohibitive energetic constraint. Upon release of pressure, the weaker internal nitrogen-nitrogen bonds are easier to break, allowing a rearrangement towards shorter chains, a process terminated by the formation of hydrazine that only incorporates a stronger terminal $H_2N-NH_2$ bond. This decomposition hypothesis is experimentally supported: first, upon pressure release starting from 50 GPa, the IR and Raman spectra are sharpening, which might indicate gradual transformation of longer chains to shorter chains. Furthermore, below 10 GPa hydrazine univocally dominates in the IR and Raman spectra (Figs 3b,4, Fig. S4,S7). Second, the Raman $H_2$ vibron intensity shows a decrease that accompanies the significant increase in the hydrazine lines in the Raman spectrum (Fig 4b, Fig. S7).

We compare the theoretically predicted spectra for mixtures of azanes (Fig. 3, Figs S8-S15) against the experimental data (Fig. 3a, Fig.S4), finding interesting points of agreement. Firstly, the spectra predict a band of N−H stretch vibrations around 3500 cm$^{-1}$. For hydrazine, this peak is well-structured, while in larger azanes, this peak broadens due to the wider variety of N-H moieties, where positioning in the chain, branching, and folding of the chains affect the individual N−H stretching modes. Compared to the higher-pressure spectra, the predicted peaks are not as broad, nor as intense, but this can be related to intermolecular interactions in the matrix. The same consideration is applied to a band around 1650 cm$^{-1}$, calculated to be the $NH_2$ deformation modes. This peak is found to be prominent in the spectra of all azanes, from N2 to N10, and broadens for more complex mixtures as a function of conformer folding, branching, and chain length. Finally, there are two broad peaks, 1500-700 cm$^{-1}$, and below 700 cm$^{-1}$, that match reasonably well with the observed spectrum. These modes are not easily assigned to specific motions as they often involve larger skeletal vibrations. Cyclic azanes, $N_nH_n$, were also examined. These, however, lack the distinct $NH_2$ deformational peak around 1650 cm$^{-1}$, and can thus be only a minor fraction of the products.

The formation of single-bonded hydrogen-nitrogen compounds was shown experimentally to occur at room temperature for pressures of about 35 GPa, opening novel ways to synthesize these high-energy materials. Compared to other high-energy materials, such as polymeric nitrogen, where polymer formation occurred at 150 GPa the needed pressures for synthesis are significantly lower. The pressure of polymerization might be further dramatically reduced as formation of the oligomer chains is energetically favorable at pressures as low as 2 GPa (Fig. 5c) according to the *ab initio* calculations and even at zero pressure according to the quantum chemical calculations. Increase of temperature is one of the ways to overcome the kinetic barrier of the reaction, and indeed this reduces the pressure of polymerization (Fig. 2). However a problem associated with high temperatures is formation of ammonia. Even at room temperature, we observed a small area of ammonia at the edge of the steel gasket (which is a catalyst) in few cases. The problem of separation of the N-H oligomers and ammonia and finding proper catalysts for the high temperatures remains to be solved. We believe that another method – ultraviolet illumination – might be effective for practical synthesis of the N-H oligomers at room temperatures and atmospheric or low pressures. UV radiation, for instance, from excimer laser with wavelength of 193 nm (~6.4 eV) can excite $N_2$ and $H_2$ molecules through one- or two-photon absorption to higher energetic states or break their bonds. Practical implementation of synthesis seems feasible with a complexity comparable to e.g. the Haber-Bosch process for synthesis of ammonia.

**Acknowledgments**. Support provided by the Max Planck Society, the DFG under grant 539-1-2, and the European Research Council under the 2010-Advanced Grant 267777 is gratefully acknowledged. M.E. and I.T. appreciate the valuable comments and support of Prof. M. O. Andreae and Dr J. Williams. Y. M. and H. L. are supported by the Natural Science Foundation of China under grants 11025418 and 91022029 and H. W. was supported by the 2012 Changjiang Scholars Program China Scholarship Council. Assistance in X-ray-diffraction measurements at beams ID-27 at ESRF and the Extreme Conditions Beamline at PETRA III, DESY is greatly appreciated. 'LV is supported by the Max Planck Graduate Center with the Johannes Gutenberg-Universität Mainz (MPGC).

**Author contributions**: M.E., H. W., and I. T. contributed equally to this paper. M.E. proposed the research and wrote the manuscript. H. W., I.T. and M.E. performed the experiments and contributed to data interpretation and writing the paper. Y. M, H. L., and L. V. performed theoretical calculations and contributed to the data interpretation and writing the manuscript.

## Methods

We filled the diamond anvil cell with nitrogen/hydrogen mixtures with the aid of a gas loader at pressures of ~1500 bar. Typically we used gasket made of T301 steel. We checked if this material containing ~70% of Fe, 17% of Cr and 7% of Ni can act as a catalyst for the $N_2$:$H_2$ mixture. For that we used an insert made of NaCl and cleaned surface of diamonds from the rest of the gasket material. We obtained the same results as with the metallic gasket at room temperature.

We used type IIa synthetic diamond anvils for IR studies and low luminescence Ia diamonds for Raman studies. The Raman spectrometer was equipped with a nitrogen-cooled CCD, notch filters, and edge filters. The 632.8 nm line of a He-Ne laser, and the 647.1 nm and 676.4 nm lines of a krypton laser were used to excite the Raman spectra. Low-temperature measurements were performed in an optical cryostat. The pressure was determined from the shift in the high-frequency edge of the Raman spectrum recorded from the stressed tip of the diamond anvil (21) or with a ruby gauge (22). The diamond anvil cell was heated with the aid of an external heater; the monitored pressure did not change appreciably.

Quantum chemical calculations were performed at the DFT level of theory, using the M06-2X functional (23) in conjunction with the aug-cc-pVDZ basis set. This level of theory is expected to be sufficiently accurate to analyze the trends in properties across the chain length; sample higher level calculations for hydrazine were conducted but the differences were found to be not significant for our current purpose. All calculations were performed using the Gaussian-09 program suite.

In the present study, the metadynamics method was applied (18, 19) with the projector augmented plane-wave (PAW) method (24), as implemented in the Vienna *ab initio* Simulation Package (VASP) code (25). A PAW potential with a Perdew-Burke-Ernzerhof (26) exchange-correlation functional was adopted. The simulation cells were constructed by using 18 nitrogen and 72 hydrogen molecules, and the Brillouin zone was sampled with Γ-point approximation. The canonical (*NVT*) ensemble was used for molecular dynamics runs. Each metastep of the metadynamics simulations comprised 600 time steps of 1.0 fs. Extensive metadynamics simulations with typically 100 metasteps for each simulation were conducted at pressures and temperatures of 30–60 GPa and 300–500 K, respectively. The width and height of the Gaussian bias potentials were $\delta = 30$ (kbar Å$^3$)$^{1/2}$ and $W = 900$ kbar Å$^3$, respectively. The metadynamics method (18, 19) is able to overcome barriers and hence can explore a broad range of candidate structures at finite temperatures. Successful applications of the method include several examples of reconstructive structural transitions (27-29).

A plane wave energy cutoff of 600 eV was employed for the underlying *ab initio* structural relaxations. The k-point sampling of 4×3×4 for N$_2$H$_8$, 7×7×7 for N$_2$ (*R*-3*c*) and 9×9×10 for H$_2$ (*P*6$_3$/*m*), respectively, were used to ensure that all the enthalpy calculations are well converged.

**Figures**

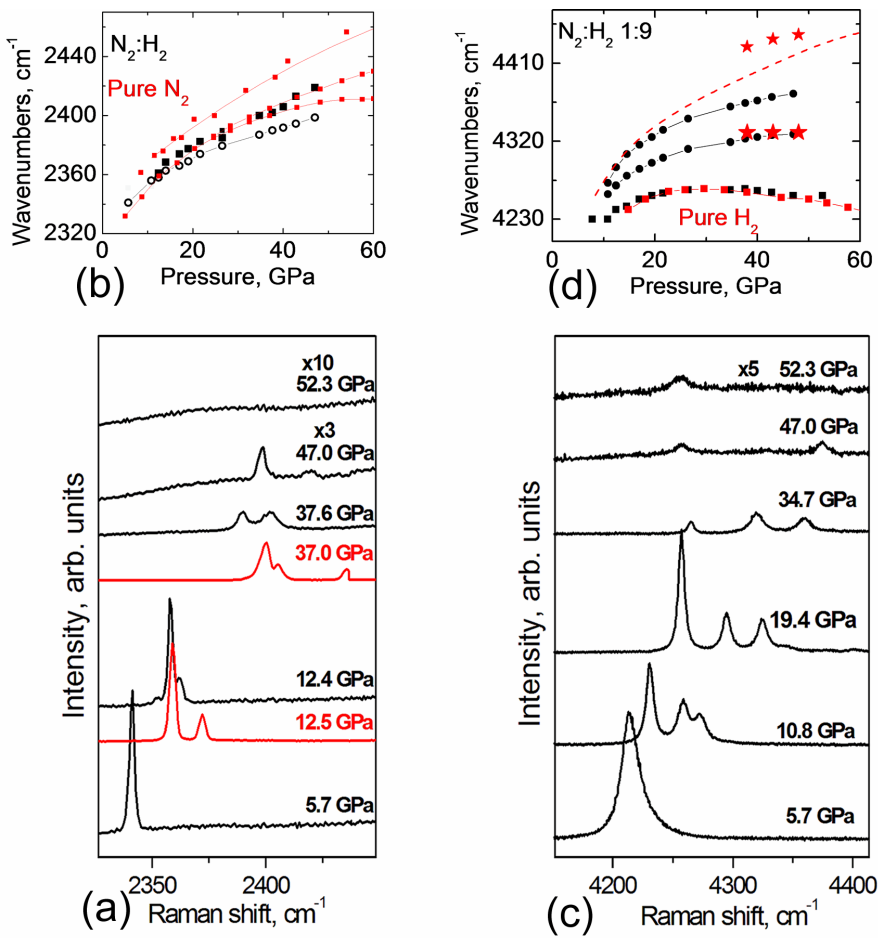

Fig. 1. Nitrogen and hydrogen Raman spectra of $N_2$:$H_2$ mixtures at high pressures. (a) Raman vibron of nitrogen in the $N_2$:$H_2$ 1:9 mixture. The spectra are shifted vertically for better comparison. Note that Raman signals at pressures of 47 and 52.3 GPa are very weak and hence amplified three- and tenfold, respectively. Red lines are the spectra of pure nitrogen. (b) Pressure dependence of the Raman vibron of nitrogen in mixtures compared to that in pure nitrogen (red points and lines). (c) Evolution of hydrogen vibrons with pressure in the 1:9 $N_2$:$H_2$ mixture. (d) Comparison of Raman and IR absorption (red stars) spectra of the 1:4 $N_2$:$H_2$ mixture with the IR vibron of molecular hydrogen (red dashed line from a previous paper (22)).

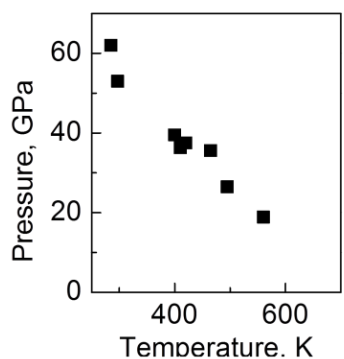

Fig. 2. The pressure-temperature conditions for the chemical transformation of the 1:1 $N_2$:$H_2$ mixture. Note that below room temperature the pressure needed for transformation significantly increases. At even colder temperatures of 200 and 250 K, we observed no chemical transformation even at pressures as high as 65 GPa. Higher temperatures substantially decrease the pressure of transformation.

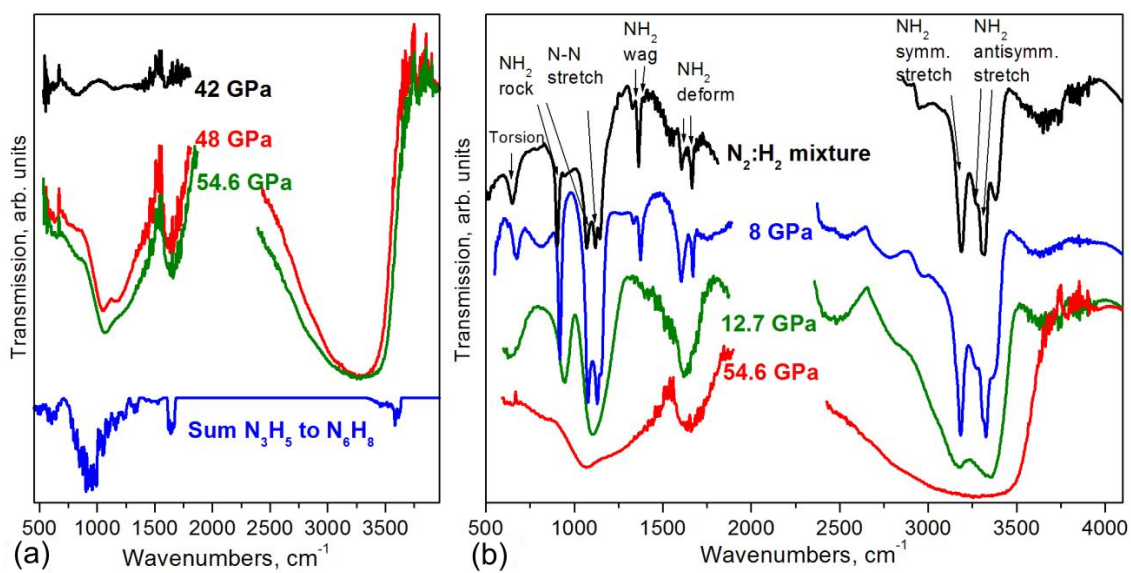

Fig. 3. Infrared spectra of $N_2:H_2$ mixtures. (a) Infrared absorption spectra of $N_2:H_2$ 1:19 mixture with increase of pressure at 300 K. A broad absorption band around 3300 cm$^{-1}$ appeared at 38 GPa (or 35 GPa in other runs with different mixtures). The oscillations in this spectrum are due to interference of light between parallel diamond anvil culets. The spectrum below (blue line) is the calculated sum absorption of oligomers from N3 to N6 chains in the gas phase. The peaks are sharp and narrow due to the absence of broadening interactions with the bulk material. Gas phase spectra of individual molecules as well as details of the calculations are presented in Figs S8-S15. (b) IR spectra of the 1:19 $N_2:H_2$ mixture at a releasing pressure after polymerization at 300 K. The spectra did not change qualitatively with pressure down to ~10 GPa. Below this pressure the spectra are identical to those of hydrazine (Fig. S4) and assigned according to Ref.(17). Note that the spectra for the 1:1 $N_2:H_2$ mixture compared to the spectra of pure hydrazine are shown in Fig. S4.

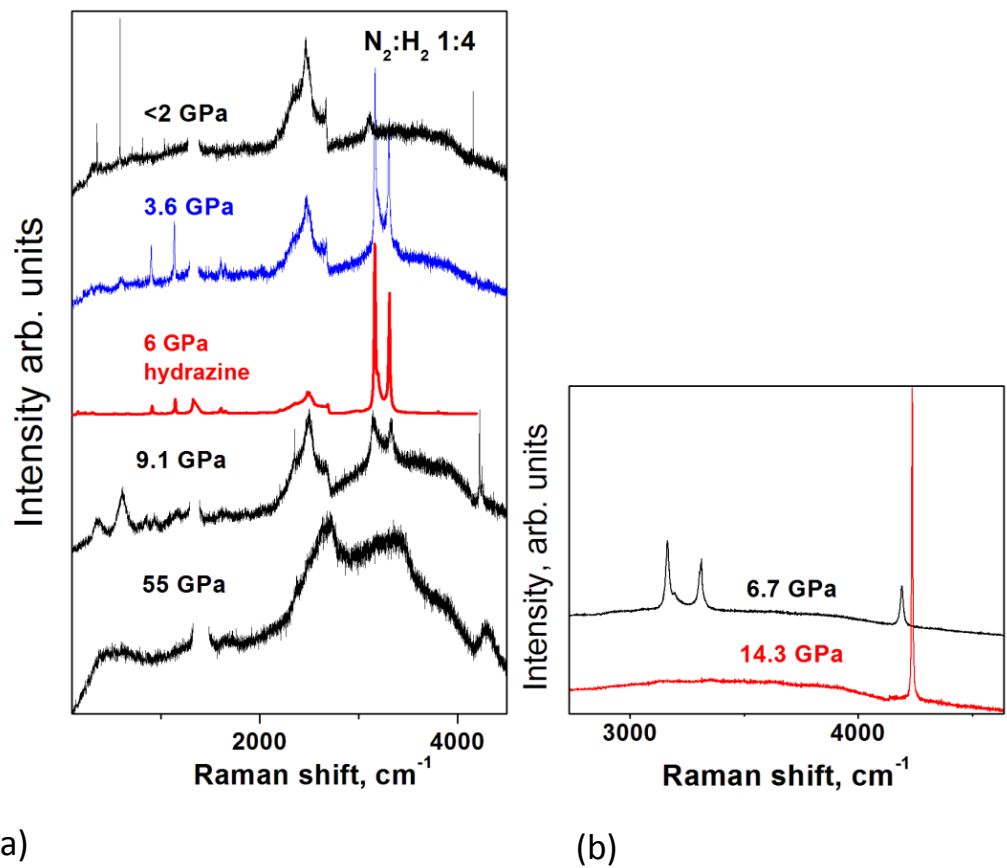

Fig. 4. (a) Raman spectra of 1:4 $N_2$:$H_2$ mixture at the releasing pressure after reaction. At the highest pressures the Raman spectra reveal only broad bands around 3200 cm$^{-1}$ and 4200 cm$^{-1}$. At pressures below ~10 GPa sharp peaks developed from these bands, identified as hydrazine(17) and hydrogen(22). The increase in the intensity of the hydrazine peaks is accompanied by a decrease on the intensity of hydrogen vibrons as shown in detail in (b) for 1:19 $N_2$:$H_2$ mixture. Hydrazine decomposes producing hydrogen and ammonia by decreasing pressure below 2 GPa before opening the cell as seen in the sharp characteristic spectra.

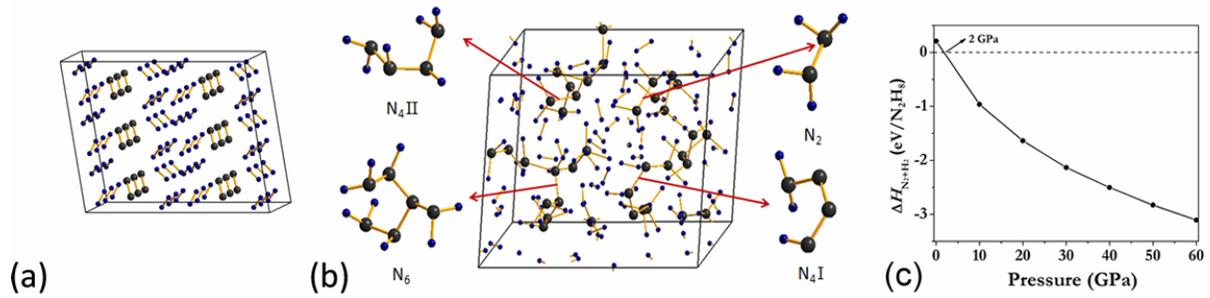

Fig. 5. High pressure phase of the 1:4 $N_2$:$H_2$ mixture as predicted theoretically using the metadynamics method. The initial (a) and predicted (b) structures of $N_{36}H_{144}$ at 60 GPa and 300 K are displayed. Large and small spheres denote nitrogen and hydrogen atoms, respectively. (b) N2, N4-II, N4-I and N6 polymeric units created at 60 GPa are in the following ratio: 7, 3, 1, 1 correspondently. (c) Calculated formation enthalpy of N-H compound with respect to the elemental decomposition into solid hydrogen ($P6_3/m$) and nitrogen ($R$-$3c$) at 0 K and high pressures.

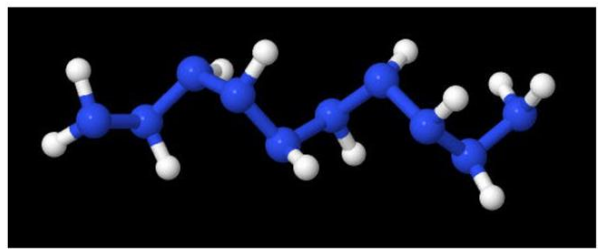

Fig. 6: Ball and stick representation of $N_{10}H_{12}$, illustrating the low-energy spiral backbone structure.

# References

1. Mailhiot C, Yang LH, & McMahan AK (1992) Polymeric nitrogen. *Phys. Rev. B* 46:14419-14435.
2. Klapotke T (2007) New nitrogen-rich high explosives. in *Structure and Bonding*, ed Klapotke T ( SPRINGER), pp 85-121.
3. Eremets MI, Gavriliuk AG, Trojan IA, Dzivenko DA, & Boehler R (2004) Single-bonded cubic form of nitrogen. *Nature Materials* 3:558-563.
4. Eremets MI, Hemley RJ, Mao HK, & Gregoryanz E (2001) Semiconducting non-molecular nitrogen up to 240 GPa and its low-pressure stability. *Nature* 411:170-174.
5. Boates B & Bonev SA (2009) First-Order Liquid-Liquid Phase Transition in Compressed Nitrogen. *Phys. Rev. Lett.* 102: 015701.
6. Boates B & Bonev SA (2011) Electronic and structural properties of dense liquid and amorphous nitrogen. *Phys. Rev. B* 83:174114.
7. Mattson WD, Sanchez-Portal D, Chiesa S, & Martin RM (2004) Prediction of new phases of nitrogen at high pressure from first-principles simulations. *Phys. Rev. Lett.* 93:125501.
8. Christe KO, Wilson WW, Sheehy JA, & Boatz JA (1999) N5: A Novel Homoleptic Polynitrogen Ion as a High Energy Density Material. *Angew. Chem. Int. Ed.* 38(13/14):2004-2009.
9. Li Y-C*, et al.* (2010) 1,1′-Azobis-1,2,3-triazole: A High-Nitrogen Compound with Stable N8 Structure and Photochromism. *J. Am. Chem. Soc.* 132:12172–12173.
10. Klapötke TM & Piercey DG (2011) 1,1′-Azobis(tetrazole): A Highly Energetic Nitrogen-Rich Compound with a N10 Chain. *Inorg. Chem.* 50: 2732–2734.
11. Bartlett RJ (2000) Exploding the mysteries of nitrogen. *Chemistry & Industry* 4: 140-143.
12. Hu A & Zhang F (2011) A hydronitrogen solid: high pressure ab initio evolutionary structure searches. *J. Phys.: Condens. Matter* 23 022203.
13. Medvedev SA*, et al.* (2011) Pressure induced polymorphism in ammonium azide (NH4N3). *Chemical Physics* 386:41.
14. Galtsov NN, Prokhvatilov AI, & Strzhemechny MA (2007) Structure of quench condensed nH2–N2 binary alloys: isotope effect. *Low Temper. Phys.* 33:499.
15. Ciezak JA, Jenkins TA, & Hemley RJ (2009) Optical and Raman microspectroscopy of nitrogen and hydrogen mixtures at high pressures. in *CP1195, Shock Compression of Condensed Matter - 2009*, ed M.L.Elert WTB, M.D.Furnish, W.W.Anderson, and W.G.Proud (American Institute of Physics), p 1291.
16. Kim M & Yoo C-S (2011) Highly repulsive interaction in novel inclusion D2–N2 compound at high pressure: Raman and x-ray evidence. *J. Chem. Phys.* 134:044519.
17. Durig JR & Zheng C (2002) On the vibrational spectra and structural parameters of hydrazine and some methyl substituted hydrazines. *Vibr. Spectr.* 30:59.
18. Martonak R, Laio A, & Parrinello M (2003) Predicting Crystal Structures: The Parrinello-Rahman Method Revisited  *Phys. Rev. Lett.* 90:075503.
19. Martonak R, Donadio D, Oganov A, & Parrinello M (2006) Crystal structure transformations in SiO2 from classical and ab initio metadynamics. *NATURE MATERIALS*  5:623-626
20. Bartlett RJ & Tobita M (Predicted Structures and Spectroscopic Characteristics of Hydrazine, Lithium-substituted Hydrazine and Their Higher Derivatives. *available on* [http://www.clas.ufl.edu/users/rodbartl](http://www.clas.ufl.edu/users/rodbartl)
21. Eremets MI (2003) Megabar high-pressure cells for Raman measurements. *J. Raman Spectroscopy* 34:515–518.
22. Mao HK & Hemley RJ (1994) Ultrahigh pressure transitions in solid hydrogen. *Rev. Mod. Phys.* 66:671-692.

23. Y. Zhao & Truhlar DG (2008) The M06 suite of density functionals for main group thermochemistry, thermochemical kinetics, noncovalent interactions, excited states, and transition elements: two new functionals and systematic testing of four M06-class functionals and 12 other functionals  *THEORETICAL CHEMISTRY ACCOUNTS*  120: 215-241
24. Blochl PE (1994) Projector augmented-wave method. *Phys. Rev. B* 50:17953.
25. Kresse G & Furthmüller J (1996) Efficient iterative schemes for ab initio total-energy calculations using a plane-wave basis set. *Phys. Rev. B* 54(16):11169.
26. Perdew JP, Burke K, & Ernzerhof M (1996) Generalized Gradient Approximation Made Simple. *Phys. Rev. Lett.* 77 3865
27. Donadio D, Martonak R, Raiteri P, & Parrinello M (2008) Influence of Temperature and Anisotropic Pressure on the Phase Transitions in α-Cristobalite. *Phys. Rev. Lett.* 100:165502.
28. Sun Jea (2009) High-pressure polymeric phases of carbon dioxide. *Proc. Natl. Acad. Sci. U.S.A.* 106:6077.
29. Behler J, Martonak R, D D, & Parrinello M (2008) Metadynamics Simulations of the High-Pressure Phases of Silicon Employing a High-Dimensional Neural Network Potential. *Phys. Rev. Lett.* 100:185501.

*Supplementary information*

# Nitrogen backbone oligomers

[1,2]H. Wang, [1]M. I. Eremets*, [1,3]I. Troyan, [2]H. Liu, [2]Y. Ma*, [1]L. Vereecken

*[1]Max Planck Institute for Chemistry, Biogeochemistry Department, PO Box 3060, 55020 Mainz, Germany*

*[2]State Key Lab of Superhard Materials, Jilin University, Changchun 130012, P. R. China*

*[3]Institute of Crystallography, Russian Academy of Sciences, Leninsky pr. 59, Moscow 119333, Russia*

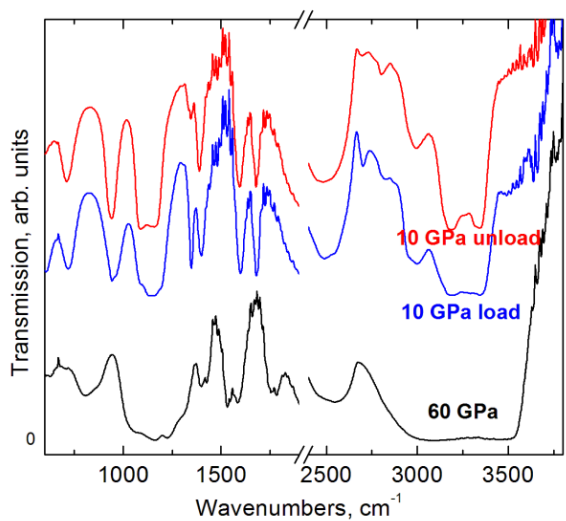

Fig. S1. Infrared spectra of hydrazine at selected pressures. The spectra gradually change with pressure and reversibly return back at releasing pressure. Spectra at intermediate pressures were omitted for clarity.

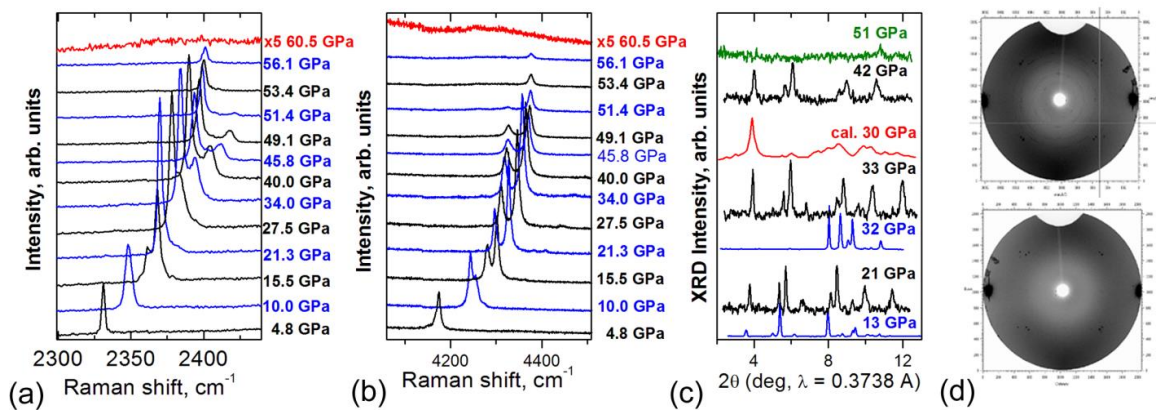

Fig. S2. Evolution of nitrogen (a) and hydrogen (b) vibron for the 1:4 $N_2$:$H_2$ mixture for increasing pressures. Note that the intensity of the 60.5 GPa spectrum was multiplied by five. (c) Experimental X-ray diffraction patterns of the 1:4 $N_2$:$H_2$ mixture (black lines) and pure nitrogen (blue lines) at different pressures. The diffraction pattern calculated from the predicted structure (see text) at 30 GPa is shown by the red curve. (d) Original images of power diffractions at 42 GPa (above) and 51 GPa (below) taken with oscillations E10°.

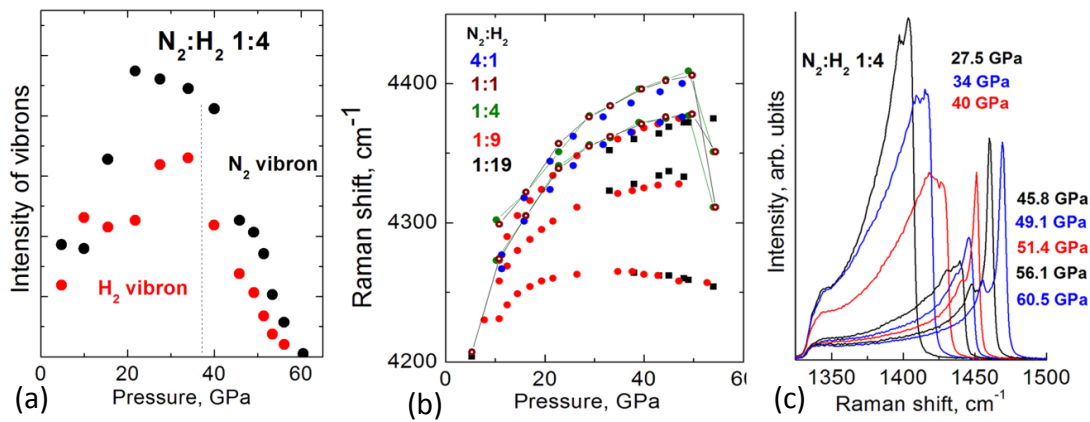

Fig. S3. Vibron changes at the polymerization. (a) The intensities of the hydrogen and nitrogen vibron substantially decrease at pressures above 35 GPa. (b) Pressure dependence of the hydrogen vibron. Note that there is a significant decrease in frequency at pressures of ~50 GPa. In the same pressure range of 40–50 GPa, the shape of the Raman signal from the stressed tip of the diamond anvils also changes significantly (c). Both the reduction in the vibron frequency and the sharpening of diamond edge indicate a concomitant volume (and pressure) decrease during polymerization.

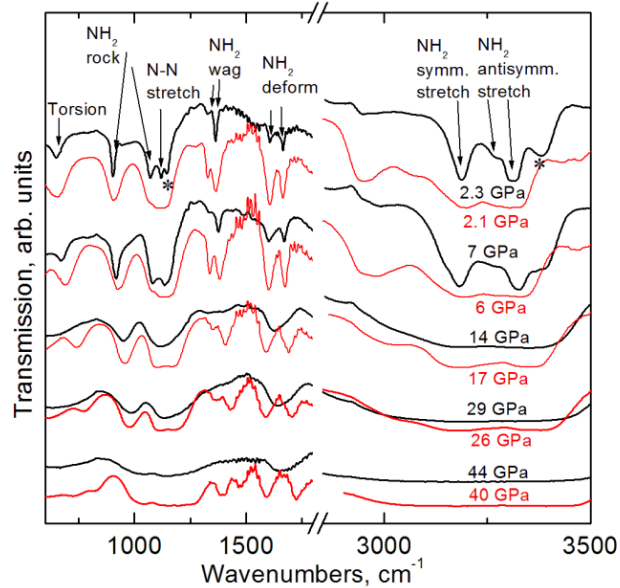

Fig. S4. IR spectra of the 1:1 $N_2$:$H_2$ mixture compared with pure hydrazine. At a releasing pressure of 55 GPa after polymerization at 300 K. The spectra of the product (black lines, pure hydrazine – red lines) did not change significantly with pressure down to 14 GPa. Below this pressure the spectra become very close to hydrazine. The only exceptions are lines marked with asterisks – they belong to ammonia. Ammonia appears in some runs, apparently the metallic gasket surrounding the sample catalyses this synthesis. Hydrazine absorption bands are assigned according to Ref. 17 of the main text.

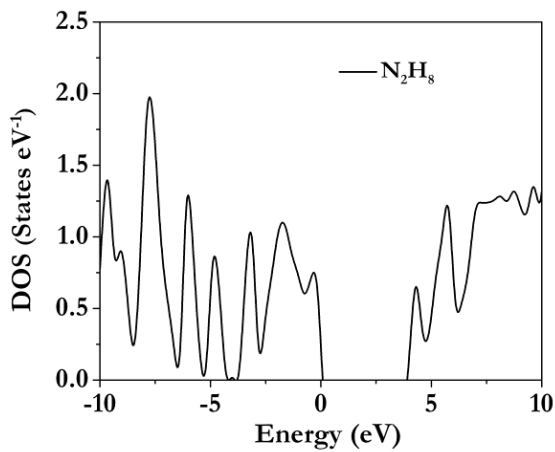

Fig. S5. The calculated total electronic density of states for N-H compound at 60 GPa is shown. The Fermi energy is set at zero. The band gap is ~3.7 eV.

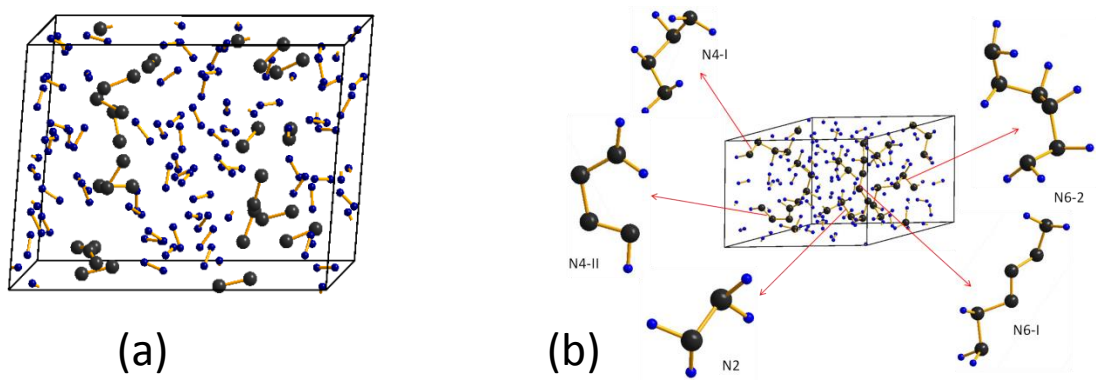

Fig. S6. Metadynamics theoretical prediction of high pressure structures. Predicted structure at (a) 30 GPa and 300 K and (b) 60 GPa at 500 K. The ratio of the polymeric units N2, N4I, N4II, N6I, N6II is: 4:3:1:1:1, correspondently.

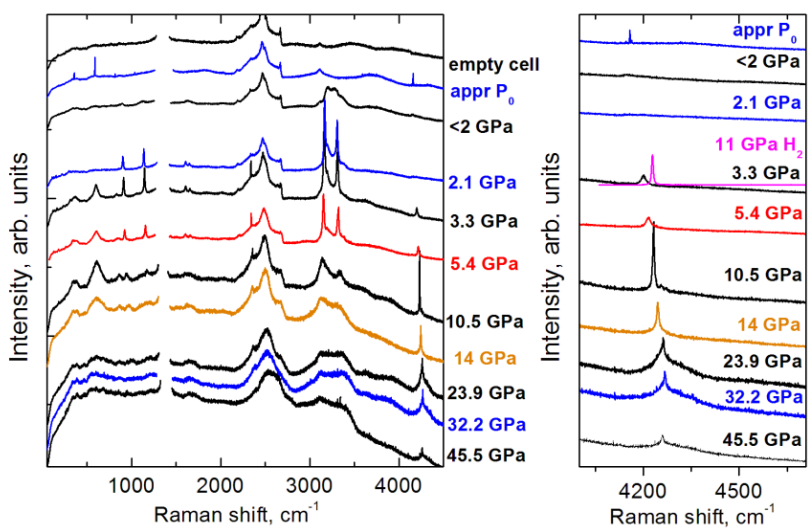

Fig. S7. Raman spectra of the 1:4 $N_2$:$H_2$ mixture at the releasing pressure. Evolution of the spectra (a) are similar to the run shown in Fig. 4a. Detailed changes of Raman spectra of hydrogen are shown in (b). At a releasing pressure below 10 GPa, hydrazine appears. Hydrogen remains inside hydrazine as its vibron is much broader than that for free hydrogen (magenta line). At the lowest pressures (not controlled) before opening the cell, hydrazine decomposes, and free hydrogen is released. Note, that molecular hydrogen presents in significant quantities in the oligomer as follows from a band at ~4300 cm$^{-1}$ in the Raman spectra. A large broadening of this peak indicates that $H_2$ molecules interact with neighboring disordered oligomers. This situation is in contrast with the nitrogen crystal (Fig. S2b) where $H_2$ molecules reveal sharp pronounced vibrons.

# Theoretical calculations on $NH_2$-$(NH)_n$-$NH_2$

## A. Quantum chemical methodology

Density functional calculations were performed to explore the potential energy surfaces of hydrazine and $NH_2$-$(NH)_n$-$NH_2$ oligomers (azanes; azoalkanes) up to $N_{10}H_{12}$, determining geometries, relative energies, and vibrational frequencies. The reasons for choosing these oligomers is discussed in the section on IR vibrational spectra. The quantum chemical calculations were performed at the DFT level of theory, using the M06-2X functional (*1*) in conjunction with the aug-cc-pVDZ basis set. This level of theory is expected to be sufficiently accurate to analyze the trends in properties across the chain length. Some of the vibrational analysis are expected to be affected by complex features of the potential energy surface, e.g. for hydrazine (*2–5*). We do not aim to accurately reproduce these subtleties, which can induce errors in the fundamental frequencies of tens of cm$^{-1}$, but rather to examine how the vibrational modes vary with respect to chain length within the same level of approximation. Sample higher level calculations for hydrazine were conducted but the differences were found to be not significant for our current purpose.

The conformer space for smaller hydronitrogen chains, up to 6 nitrogen atoms in the backbone, was sampled exhaustively to determine the lowest energy conformers. The geometric patterns in these conformers were then generalized to longer chains up to 10 N-atoms, to obtain a good estimate of their optimal geometries. While this approach should give fairly good results, more stable conformers might still exist for these longer oligomers. For the current set of exploratory calculations only unbranched azane chains were examined, with the exception of a cyclic structure. Also, only isolated molecules at zero pressure are examined theoretically at this time. As discussed below, this will impact some parts of the vibrational spectra.

All calculations were performed using the Gaussian-09 program suite (*6*). The ball-and-stick figures were prepared with Jmol.(*7*)

## B. Computational results

### B.1. Geometries

For hydrocarbon chains, the most stable conform is known to be the linear backbone, with all carbons in a plane. In contrast, the $NH_2$-$(NH)_n$-$NH_2$ chains show a preference for an N-N-N-N dihedral angle of about 76 degrees, causing the backbone to adopt a spiral shape (see Figure 3 in the main text); the left- and right-handed spirals were found to be enantiomers with identical properties. The most stable conformers also have all -NH-hydrogen atoms with H-N(N)N dihedral angles of the same sign. This spiral shape is clearly induced by the differences in interaction between the N-H bonds versus the lone pair on nitrogen, which is also the driving force in the hydrazine geometry. The inner N-N and N-H bond lengths of about 1.41 Å and 1.02 Å, respectively, are near-identical to those at the terminating N-NH$_2$ moiety (1.42 Å and 1.01 Å).

For cyclic N$_n$H$_n$ compounds, no extensive conformer analysis was performed. For N$_6$H$_6$, a chair form was found to be the most stable structure, in agreement with earlier work (*8, 9*). We refer to these latter publications, and the references therein, for an overview of the limited available work on cyclic azanes.

### B.2. Thermodynamic stability

All NH$_2$-(NH)$_n$-NH$_2$ molecules were found to be comparatively stable energetic energy minima. Compared to the constituting elements in their elemental form, N$_2$ and H$_2$, the enthalpy of formation raises by about 20 kcal mol$^{-1}$ for each added -NH- monomer (see *Table S1*). The successive endothermicities in elongating the chain are therefore near-identical as those for forming hydrazine. The formation enthalpy for the energetically most stable conformer, in all cases showing the spiral backbone discussed above, shows a slight dependence on the chain length, with a maximum at N$_5$H$_7$, and a modest decrease in formation enthalpy at 0 K after that (see *Table S1*). While this might suggest that, once past a certain chain length, elongating an existing chain further might be energetically more favorable than forming shorter chains, the energy differences are too small to be significant at the current level of theory. The peak in enthalpy increase around N$_5$H$_7$ is related to the length of the spiral, where 5 to 6 nitrogen atoms is the minimum length to form a full revolution of the spiral backbone.

### B.3. Formation and decomposition pathways

The chemical pathways in forming, elongating, and breaking the chains seems to involve a complex set of radical reactions. Their study is well beyond the scope of this work, and the reader is referred to the literature on the analogous hydrazine (catalytic) combustion for possible reaction candidates. Unimolecular H-elimination is easier for the inner hydrogen, with a bond strength of 68 kcal mol$^{-1}$ compared to 74 for the -NH$_2$ hydrogens. H$_2$-elimination at the terminal -NH$_2$ group requires 69 kcal mol$^{-1}$. Breaking of the nitrogen chain requires 50 kcal mol$^{-1}$ at the chain end, forming ·NH$_2$ radicals, but only 30 kcal mol$^{-1}$ for the inner N-N bonds, forming a pair of -N•H + •NH- radicals. This latter pathway is significantly below the unimolecular decomposition channels for hydrazine and its alkylated forms, the lowest of which involves breaking of the N-N bond with a barrier of at least 60 kcal mol$^{-1}$.

Table S1: Formation enthalpy (kcal mol$^{-1}$) at zero Kelvin, and relative energies (kcal mol$^{-1}$) upon elongating the chain by one NH monomer.

| Compound | $\Delta H(0K)$ | $E_{rel}(N_xH_x$ to $N_{x-1}H_{x-1})$ |
|---|---|---|
| $N_2+H_2$ | 0.0 | |
| $N_2H_4$ | 19.1 | 19.1 |
| $N_3H_5$ | 39.0 | 19.9 |
| $N_4H_6$ | 58.8 | 19.8 |
| $N_5H_7$ | 79.43 | 20.6 |
| $N_6H_8$ | 109.1 | 19.9 |
| $N_7H_9$ [a] | 119.2 | 19.8 |
| $N_8H_{10}$ [a] | 139.0 | 19.8 |
| $N_9H_{11}$ [a] | 158.6 | 19.7 |
| $N_{10}H_{12}$ [a] | 178.2 | 19.6 |

**B.4. Infrared vibrational spectrum**

The gas phase vibrational spectra obtained theoretically are only partially comparable to those in liquid and solid phase. In particular, vibrational modes involving large skeletal movements are more hindered in the solid phase, often with significant shifts in the vibrational frequencies. Still, we can compare several spectral bands between hydrazine and the oligomers, and use the results to qualitatively support the experimentally observed spectral changes. The experimental data is expected to be obtained for a mixture of many nitrogen-hydrogen compounds ranging, of unknown chain length, branching, cyclic forms, or saturation. Comparison of the experimental IR spectra to the theoretically predicted fundamental wavenumbers for individual compound conformers is therefore not overly revealing. Below, we instead show spectra of mixtures of compounds, so the dependence of the spectrum on the mixture can be visualized. Two elementary spectra, for hydrazine (Fig. S8) and cyclic $N_6H_6$ (Fig. S9) are shown, as well as the spectra for mixtures containing a) the lowest conformer characterized for each of azanes from $N_3$ to $N_{10}$ in equal concentrations (Fig. S10), b) all conformers of $N_3H_5$ in equal proportion (Fig. S11), c) all conformers of $N_4H_6$ in equal proportion (Fig. S12), d) all conformers of $N_5H_7$ in equal proportion (Fig. S13), e) all conformers of $N_6H_8$ in equal proportion (Fig. S14) and f) all conformers for $N_3H_5$ through $N_6H_8$ with each conformer in equal concentration (Fig. S15). These synthetic spectra were generated from the vibrational modes and their IR intensities as predicted from the M06-2X/aug-cc-pVDZ calculations, assuming gaussian-function-shaped peaks with an arbitrary width at half height of about 8 $cm^{-1}$, and renormalized to the strongest absorption peak.

An important difference between the cyc-$N_6H_6$ spectrum, compared to the experimental spectrum and the spectra for the linear azanes is the absence of peaks near 1650 $cm^{-1}$, which corresponds to the $NH_2$ wagging modes. The prominence of this peak in the experimental spectrum suggests strongly that unsubstituted cyclic azanes can therefore not be the main products. As such, we exclude cyclic oligomers from our discussion below.

The NH stretch vibrations all have wavelengths above 3400 $cm^{-1}$, both for the oligomers and hydrazine. In hydrazine, the coupling between the eigenstates of the C-H vibrations of the -$NH_2$ moiety leads to their separating into a symmetric and an antisymmetric stretch. For the oligomers, the bulk of the hydrogen atoms are singletons, leading to much weaker coupling and no clear splitting of the spectrum, as observed experimentally. Furthermore, the reaction vessel is expected to contain a mixture of different oligomers lengths, branching, and degree of folding, each affecting the surroundings of the individual hydrogen atoms slightly, and hence leading to slightly different vibrational frequencies. For the longer oligomers, $N_9H_{11}$ and $N_{10}H_{12}$, we already predict vibrational modes ranging from 3424 to 3613 $cm^{-1}$ with the individual frequencies spread out more or less evenly across this band. This seems to support the broad peak around 3500 $cm^{-1}$ seen in the experiment. Oligomer branching should introduce further peak broadening, as this introduces further variations of the H-atom environments.

A second spectral band of interest are the peaks just above 1600 $cm^{-1}$. In hydrazine, these correspond to the wagging of the two $NH_2$ moieties; these modes again couple strongly into a symmetric and antisymmetric eigenfunction resulting in a split-peak spectrum. For $N_3H_5$, the predicted vibrational $NH_2$ wagging modes are already significantly less coupled, with each mode showing a large-amplitude wagging on one $NH_2$ end, and only a small amplitude wagging at the other $NH_2$ moiety. Upon further elongation of the nitrogen backbone, the two $NH_2$ wagging modes become fully localized on one or the other -$NH_2$ moiety. This decoupling of the $NH_2$ wagging mode, combined with small changes in the $NH_2$ environment by chain branching and folding, would prevent the formation of two separate peaks, but rather yield a single peak, as is observed experimentally. Again, oligomer branching should broaden this peak further.

For the vibrational spectrum of the oligomers at frequencies below 1500 cm$^{-1}$, the gas phase predictions are not expected to be accurate, as larger amplitude skeleton vibrations behave differently in solid phase, with typically a shift towards higher frequencies. This is particularly true for the broad peak in the spectra below 700 cm$^{-1}$. Rather than over-interpreting the predicted oligomer spectrum in this region, we merely note that the experimental spectra and the theoretically predicted spectra remain well comparable. A more accurate comparison of the spectra requires the application of different computational techniques explicitly calculating spectra for the high-pressure solid phase.

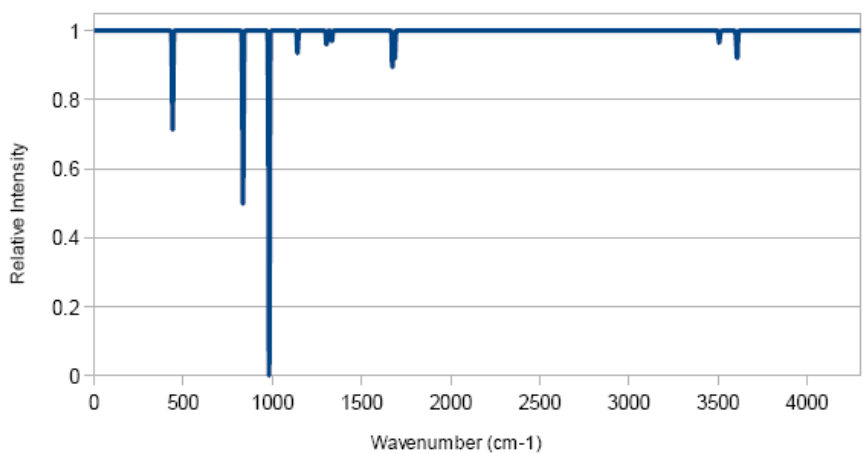

Fig. S8: Absorption spectrum of hydrazine as obtained at the M06-2X/aug-cc-pVDZ level of theory. Simulated with a peak width at half height of 8 cm$^{-1}$.

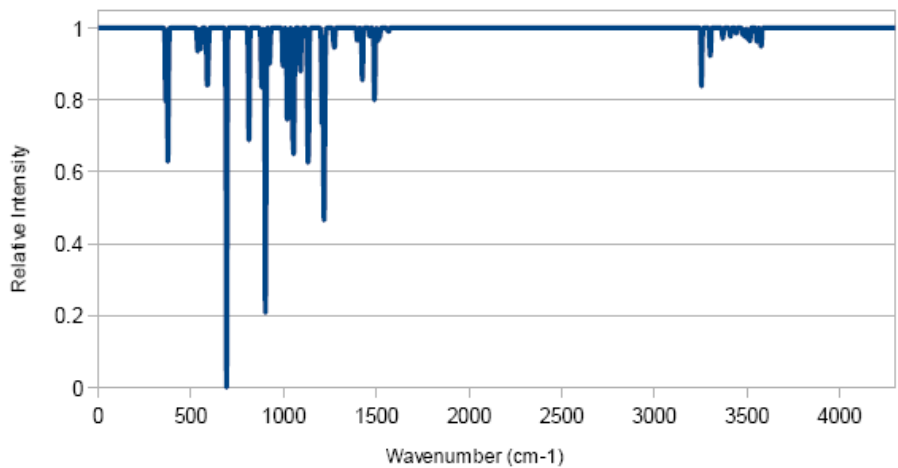

Fig. S9: Absorption spectrum of cyclic-$N_6H_6$ as obtained at the M06-2X/aug-cc-pVDZ level of theory. Simulated with a peak width at half height of 8 cm$^{-1}$.

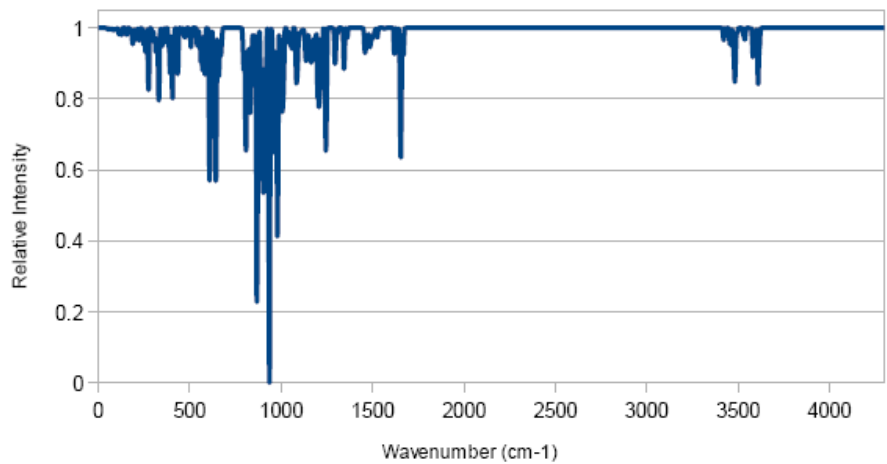

Fig. S10. Absorption spectrum of a mixture of $N_3H_5$ through $N_{10}H_{12}$, with one conformer each in equal concentration, as obtained at the M06-2X/aug-cc-pVDZ level of theory. Simulated with a peak width at half height of 8 cm$^{-1}$.

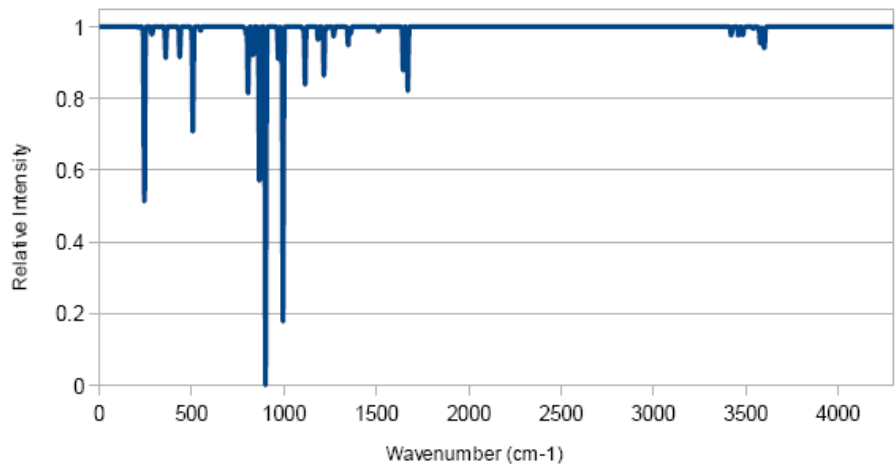

Fig. S11: Absorption spectrum of a mixture of $N_3H_5$ conformers in equal concentration, as obtained at the M06-2X/aug-cc-pVDZ level of theory. Simulated with a peak width at half height of 8 cm$^{-1}$.

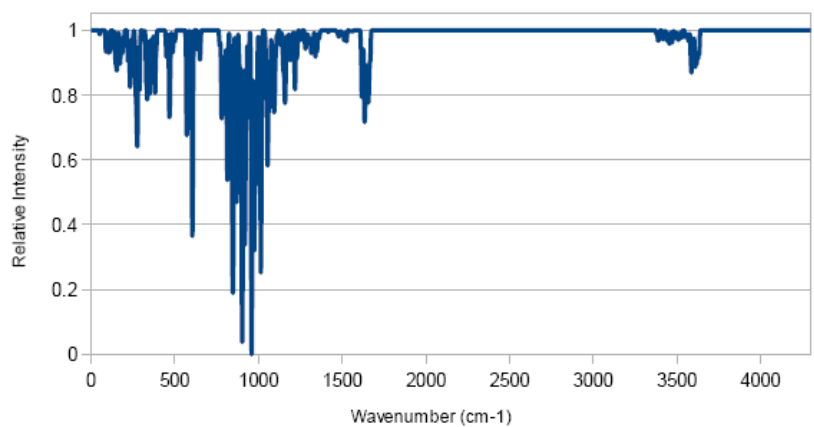

Fig. S12: Absorption spectrum of a mixture of $N_4H_6$ conformers in equal concentration, as obtained at the M06-2X/aug-cc-pVDZ level of theory. Simulated with a peak width at half height of 8 cm$^{-1}$.

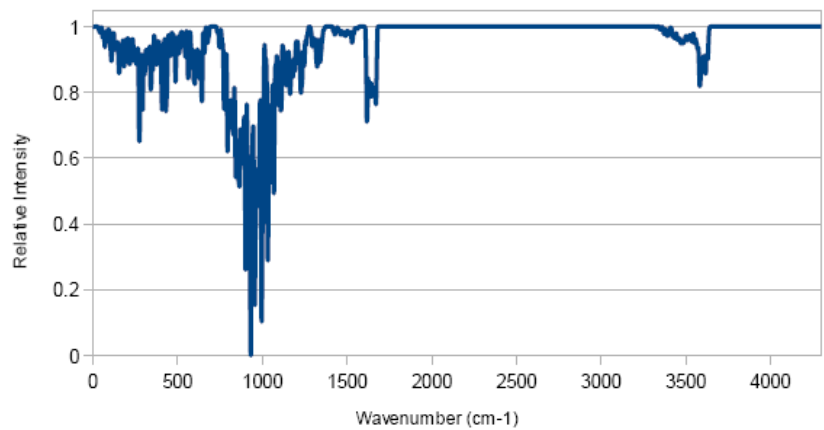

Fig. S13: Absorption spectrum of a mixture of $N_5H_7$ conformers in equal concentration, as obtained at the M06-2X/aug-cc-pVDZ level of theory. Simulated with a peak width at half height of 8 cm$^{-1}$.

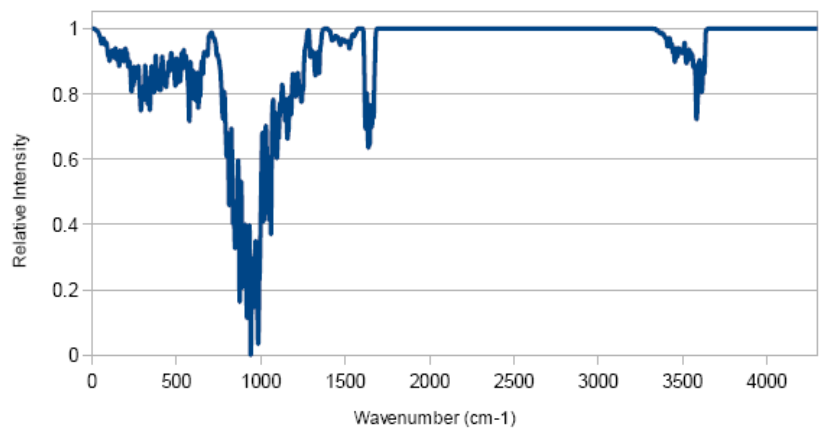

Fig. S14: Absorption spectrum of a mixture of $N_6H_8$ conformers in equal concentration, as obtained at the M06-2X/aug-cc-pVDZ level of theory. Simulated with a peak width at half height of 8 cm$^{-1}$.

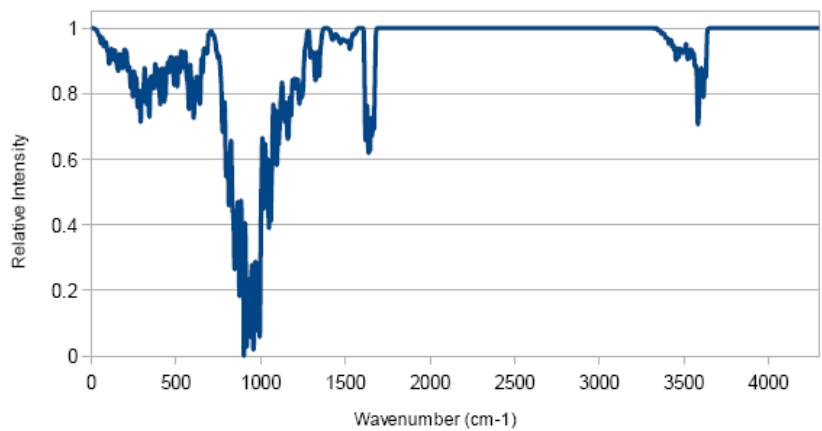

Fig. S15: Absorption spectrum of a mixture of all $N_3H_5$ through $N_6H_8$ conformers in equal concentration, as obtained at the M06-2X/aug-cc-pVDZ level of theory. Simulated with a peak width at half height of 8 cm$^{-1}$.

**References**

1. Y. Zhao, D. G. Truhlar (2008) *Theor. Chem. Accounts* **120**, 215–241

2. D. W. Ball (2001) *J. Phys. Chem. A* **105**, 465–470

3. F. B. C. Machado, O. Roberto-Neto (2002) *Chem. Phys. Lett.* **352**, 120–126

4. A. T. Kowal (2003) *J. Mol. Struct. Theochem* **625**, 71–79

5. W. Łodyga, J. Makarewicz (2012) *J. Chem. Phys.* **136**, 174301.

6. M. J. Frisch *et al.*, *Gaussian 09, Revision B.01* (Gaussian Inc., Wallington CT, 2010).

7. *Jmol: an open-source Java viewer for chemical structures in 3D* (www.jmol.org).

8. H.-Y. Wu, W.-F. Cai, L.-C. Li, A.-M. Tian, N.-B. Wong (2011) *J. Comput. Chem.* **32**, 2555–2563

9. B.-H. Xu, L.-C. Li, L. Sun, A.-M. Tian (2008) *J. Mol. Struct. Theochem* **870**, 77–82